\documentclass[
reprint,
superscriptaddress, 
amsmath,amssymb,
aps,
prl,
]{revtex4-1}

\usepackage{graphicx}% Include figure files
\usepackage{dcolumn}% Align table columns on decimal point
\usepackage{bm}% bold math
\usepackage{color}
\usepackage{hyperref}% add hypertext capabilities

\begin{document}

\preprint{APS/123-QED}

\title{Giant Rashba-type spin splitting through spin-dependent interatomic-hopping}

\author{Jisook Hong}
  \affiliation{Department of Chemistry, Pohang University of Science and Technology, Pohang 37673, Korea}
\author{Jun-Won Rhim}
  \affiliation{Max-Planck-Institut f\"ur Physik komplexer Systeme, 01187 Dresden, Germany}
\author{Inkyung Song}
  \affiliation{Center for Correlated Electron Systems, Institute for Basic Science, Seoul 08826, Korea}
\author{Changyoung Kim}
  \affiliation{Center for Correlated Electron Systems, Institute for Basic Science, Seoul 08826, Korea}
  \affiliation{Department of Physics and Astronomy, Seoul National University, Seoul 08826, Korea}
\author{Seung Ryong Park}
  \email{AbePark@inu.ac.kr}
  \affiliation{Department of Physics, Incheon National University, Incheon 22012, Korea}
\author{Ji Hoon Shim}
  \email{jhshim@postech.ac.kr}
  \affiliation{Department of Chemistry, Pohang University of Science and Technology, Pohang 37673, Korea}
  \affiliation{Department of Physics and Division of Advanced Nuclear Engineering, Pohang University of Science and Technology, Pohang 37673, Korea}

\date{\today}

\begin{abstract}
We have performed density functional theory calculation and tight binging analysis in order to investigate the mechanism for the giant Rashba-type spin splitting (RSS) observed in Bi/Ag(111). We find that local orbital angular momentum induces momentum and spin dependent charge distribution which results in spin-dependent hopping. We show that the spin-dependent interatomic-hopping in Bi/Ag(111) works as a strong effective field and induces the giant RSS, indicating that the giant RSS is driven by hopping, not by a uniform electric field. The effective field from the hopping energy difference amounts to be $\approx$18 V/\AA. This new perspective on the RSS gives us a hint for the giant RSS mechanism in general and should provide a strategy for designing new RSS materials by controlling spin-dependence of hopping energy between the neighboring atomic layers.
\end{abstract}

\maketitle

%Introuction

There has been a recent surge in the study of Rashba-type spin splitting (RSS) due to its role in the field of spintronics \cite{Manchon15,Bindel16,Zhenzhen17} as exemplified by spin field effect transistor \cite{Datta90,Koo09,Chuang15}, spin orbit torque \cite{Manchon08,Miron10} and spin to charge conversion studies \cite{Sanchez13,Lesne16,Oyarzun16}. In addition to achieving controllability of the splitting \cite{DiSante13,Kepenekian15,Volobuev17}, an important direction in the research is to increase the splitting energy \cite{Ast07,Ishizaka11}. For that reason, there have been extensive studies on the so-called giant RSS systems such as Bi/Ag(111) and Pb/Ag(111) \cite{Ast07,Bihlmayer07,Ast08,Meier08,Bian12,El-Kareh13}. 
Several proposals have been made to explain the giant RSS such as the role of in-plane potential gradient \cite{Premper07,Bentmann09,Bian13}. However, they could not provide quantitative explanation for the giant RSS. 

RSS was conventionally understood to be from an effective Zeeman coupling between the electron spin and a relativistic magnetic field for a moving electron in an electric field \cite{Bychkov84}, but it was recognized that its energy scale is too small to induce the split energy \cite{LaShell96,Petersen00}. It was recently shown that local orbital angular momentum (OAM) induces an electric polarization, and its coupling to the electric field can fully account for the scale of the energy splitting \cite{Park11,Kim13,Hong15}. Even in this picture, the electric field perpendicular to the surface or interface plays a crucial role in typical systems such as Au(111) surface states \cite{LaShell96}. However, the extraordinary giant RSS found in Bi alloy on Ag(111) challenges the role of perpendicular electric field \cite{Ast07}.

While it is evident that OAM should still play the key role by inducing the momentum dependent polarization to account for the energy scale, the electric field required for the giant RSS is unreasonably strong. A possible solution to the issue may come from an effective electric field which can be much stronger than the ordinary field from potential gradient. Our strategy is to focus on the effect of the OAM on the hopping energy and investigate it by using density functional theory (DFT) calculation as well as tight binding (TB) analysis. We find that charge configurations for different spin states vary, which result in an spin-dependent interatomic-hopping strength. We therefore argue that it is the spin-dependent interatomic-hopping that plays the role of the inversion symmetry breaking (ISB) field to induce the giant RSS in Bi/Ag(111), not the conventional electric field. Our new picture should be applicable to other systems with large RSS.

%%% Figure 1
\begin{figure*}
\includegraphics[width=\textwidth]{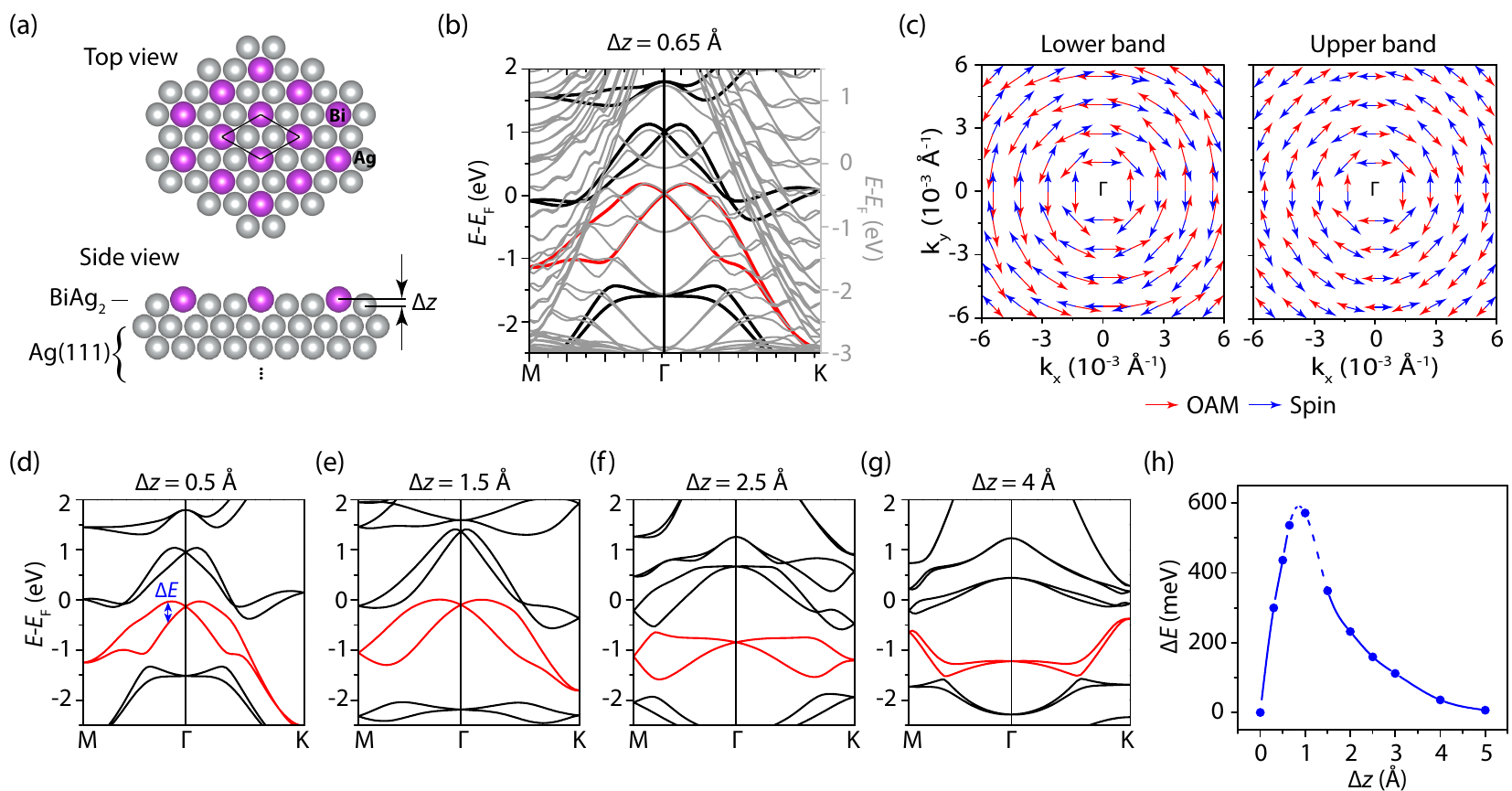}
\caption{\label{intro}
(a) Crystal structures of Bi/Ag(111) and its topmost layer BiAg$_2$. 
Figures are produced by using VESTA. \cite{Momma11} 
(b) Electronic structures of BiAg$_2$ (left axis, black and red thick lines) and Bi/Ag(111) (right axis, grey thin lines) at $\Delta z=0.65$ \AA. 
(c) Spin and OAM textures of $J\approx 1/2$ bands of BiAg$_2$ in reciprocal $\mathbf{k}$-space.
(d)-(g) Electronic structures of BiAg$_2$ for various $\Delta z$ values. Red lines denote the Bi $J\approx1/2$ bands.
(h) The size of RSS at $k=0.02$ \AA$^{-1}$ ($\Delta E$) as a function of $\Delta z$.
The maximum $\Delta E$ value occurs when $\Delta z$ is in between 0.65 and 1.5 \AA.
}
\end{figure*}

%Calculation Details

For non-collinear DFT calculations, we use  Vienna \textit{Ab initio} Simulation Package (VASP) \cite{Kresse93, Kresse94, Kresse96a, Kresse96b} and OpenMX codes \cite{Ozaki03, Ozaki04, Ozaki05}. Using VASP code with the generalized gradient approximation of Perdew-Berke-Ernzerhof (GGA-PBE) \cite{Perdew96}, we optimize the lattice constant of \textit{fcc} Ag until the internal atomic force becomes less than $10^{-8}$ eV/\AA{} and get $a=4.15$ \AA. From the lattice constant, we construct the structures of Bi/Ag(111) and BiAg$_2$, and calculate the band structures, spin and OAM. By taking Ag away from BiAg$_2$, we construct Bi triangular monolayer and calculate the electronic structures. For analysis on the orbital composition of the Rashba-split bands, we use OpenMX code with $s2p2d2$ pseudo-atomic basis orbitals and PBE potentials for Bi and Ag atoms.

For the analysis on the relation between the RSS and other physical quantities, we construct a minimal TB model based on the DFT results and analyze it within the perturbation theory.
We consider, for each spin, one $s$-orbital at each of two Ag atoms in the unit cell, and $p_x$-, $p_y$-, and $p_z$-orbitals at Bi atom. 
Hamiltonian matrix elements are obtained following the Slater-Koster's scheme \cite{Slater54}. We find parameter sets for the bond integrals that qualitatively reproduce the giant RSS  and overall band structures.

%Results

In Fig. \ref{intro}(a), we present the crystal structure of Bi/Ag(111) alloy system. It was previously noticed that the surface states are strongly localized in the top-most layer \cite{Bian13}. We confirm this from the fact that the band structures of  Bi/Ag(111) and BiAg$_2$ monolayer show almost identical giant RSS. (Fig. \ref{intro}(b)) Figure \ref{intro}(c) shows the chiral spin and OAM textures of the split bands of BiAg$_2$ which clearly reproduce the essential features of Bi/Ag(111) surface states \cite{Schirone15}. As the Rashba states are almost the same, we continue our discussion based on the BiAg$_2$ for the sake of simplicity (no bulk states present).  

It has been known that the size of RSS in BiAg$_2$ is highly sensitive to the buckling distance $\Delta z$ in Fig. \ref{intro}(a) \cite{Bihlmayer07,Bian13}. In Figs. \ref{intro}(d)-(g), we present the band structures for various $\Delta z$ values and highlight the Bi $J\approx1/2$ bands (red lines) which show the largest RSS. The splitting energy ($\Delta E$) against $\Delta z$ is depicted in Fig. \ref{intro}(h). As expected, there is no splitting when $\Delta z$ is zero since there is no ISB. As $\Delta z$ increases, $\Delta E$ initially increases very sharply. It hits the maximum at $\approx$1 \AA{} and then decreases.  When $\Delta z$ is far enough, the band structure becomes independent bands of Ag hexagonal lattice and Bi triangular monolayer.

%%% Figure 2
\begin{figure}
\includegraphics[width=\columnwidth]{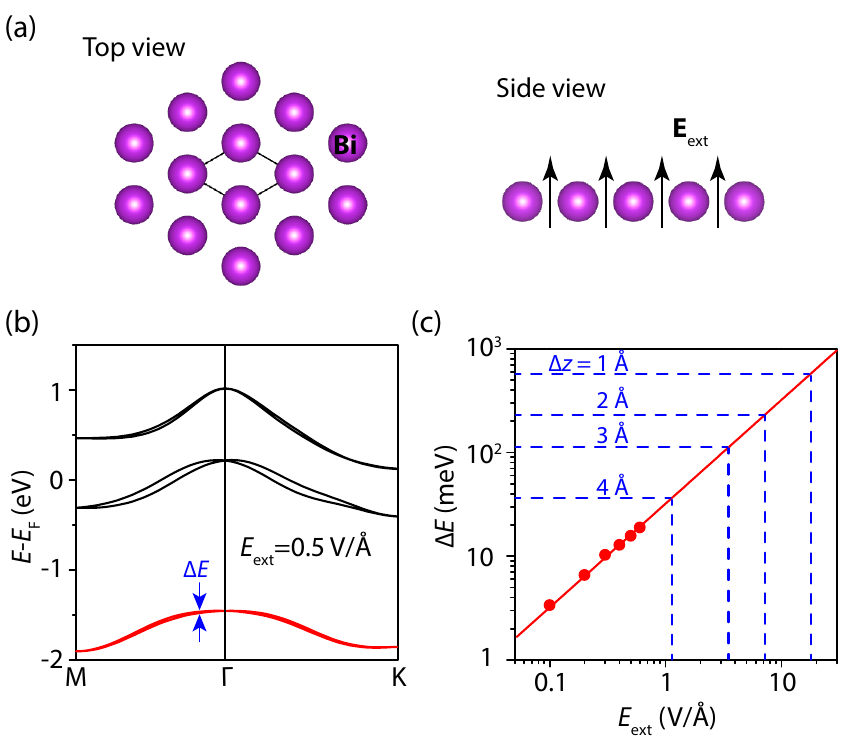}
\caption{\label{Bi_mono}
(a) Crystal structure of Bi triangular monolayer. An external electric field $E_\text{ext}$ is applied perpendicular to the Bi plane.
(b) Electronic structure of Bi monolayer under $E_\text{ext}=0.5$ V/\AA. $J\approx1/2$ bands are shown in red.
(c) The splitting energy at $k=0.02$ \AA$^{-1}$ ($\Delta E$) as a function of $E_\text{ext}$. Blue dashed lines represent the required $E_\text{ext}$ for Bi monolayer to have the same $\Delta E$ of the BiAg$_2$ $J\approx1/2$ bands, estimated at different $\Delta z$ values.
}
\end{figure}

We first check out the possibility that the giant RSS is induced by a uniform surface field. In order to study the possible surface field, we compare the $\Delta E$ of BiAg$_2$ and Bi triangular monolayer under an external field. (Fig. \ref{Bi_mono}(a)) We plot the band structure of Bi monolayer under $E_\text{ext}=0.5$ V/\AA{} in Fig. \ref{Bi_mono}(b). The pair of bands colored in red is mainly from Bi $J\approx1/2$ states and shows a RSS energy of 29 meV at $k=0.06$ \AA$^{-1}$, which is comparable to one obtained in previous studies but much smaller than the value observed in BiAg$_2$ \cite{Ast07,Koroteev04}.  In Fig. \ref{Bi_mono}(c), we plot $\Delta E$ of Bi triangular monolayer as a function of the applied field strength $E_\text{ext}$. The $\Delta E$ increases linearly with $E_\text{ext}$ as the linear fit shows (red line). Based on the extrapolation of the data, one can estimate the $E_\text{ext}$ required to split the $J\approx1/2$ bands of Bi as much as those of BiAg$_2$. A few cases for different values of $\Delta z$ are presented with blue dashed lines in the figure. One can see that a field stronger than $\approx$18 V/\AA{} is needed to make $\Delta E$ of Bi monolayer comparable to that of BiAg$_2$ with $\Delta z=1$ \AA.  The TB analysis we considered in our previous study \cite{Hong15} also shows that we need to apply $E_\text{ext} \sim V_{sp\sigma}/V_{ppz}$ (eV/\AA) $\sim$ 10 (eV/\AA) to achieve a similar $\Delta E$ ($V_{sp\sigma}$ is the bond integral between $s$- and $p$-orbitals in the $\sigma$ bond, and $V_{ppz}$ is the electric-field-induced hopping between neighboring $p_z$ orbitals in the Bi monolayer). Considering typical values of work function (a few eV) and the length scale of the surface depth (a few \AA) \cite{Petersen00}, it is unrealistic for Bi to have an effective field as strong as $\approx$18 V/\AA{} from Ag sublayer. This implies that the effective surface field is not the probable cause of the giant RSS observed in Bi/Ag(111).

%%% Figure 3
\begin{figure}
\includegraphics[width=\columnwidth]{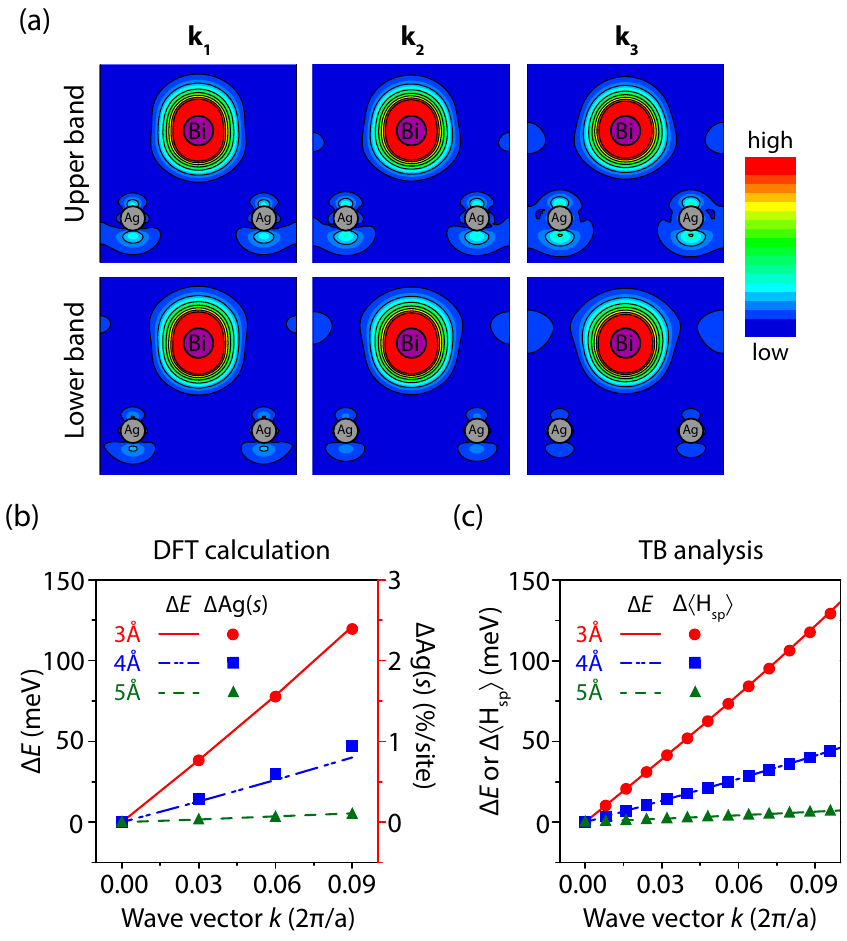}
\caption{\label{along_k}
(a) Partial charge densities of the Rashba-split bands of BiAg$_2$ with $\Delta z = 4$ \AA{} at selected $\mathbf{k}$-points of $\mathbf{k}_1=(0.06,0,0)$, $\mathbf{k}_2=(0.12,0,0)$ and $\mathbf{k}_3=(0.18,0,0)$ in unit of $2\pi/a$. We merge partial densities of six symmetrically indistinguishable $\mathbf{k}$-points.
(b) The size of splitting (left axis, lines) at wave vector $(k,0,0)$ and difference in Ag $s$-orbital contributions between the split bands (right axis, markers) from DFT calculations.
(c) The size of splitting (lines) and the difference in the averages of $s$-$p$ hopping energy between two split bands (markers) from TB calculations.
}
\end{figure}

%%% Figure 4
\begin{figure*}
\includegraphics[width=\textwidth]{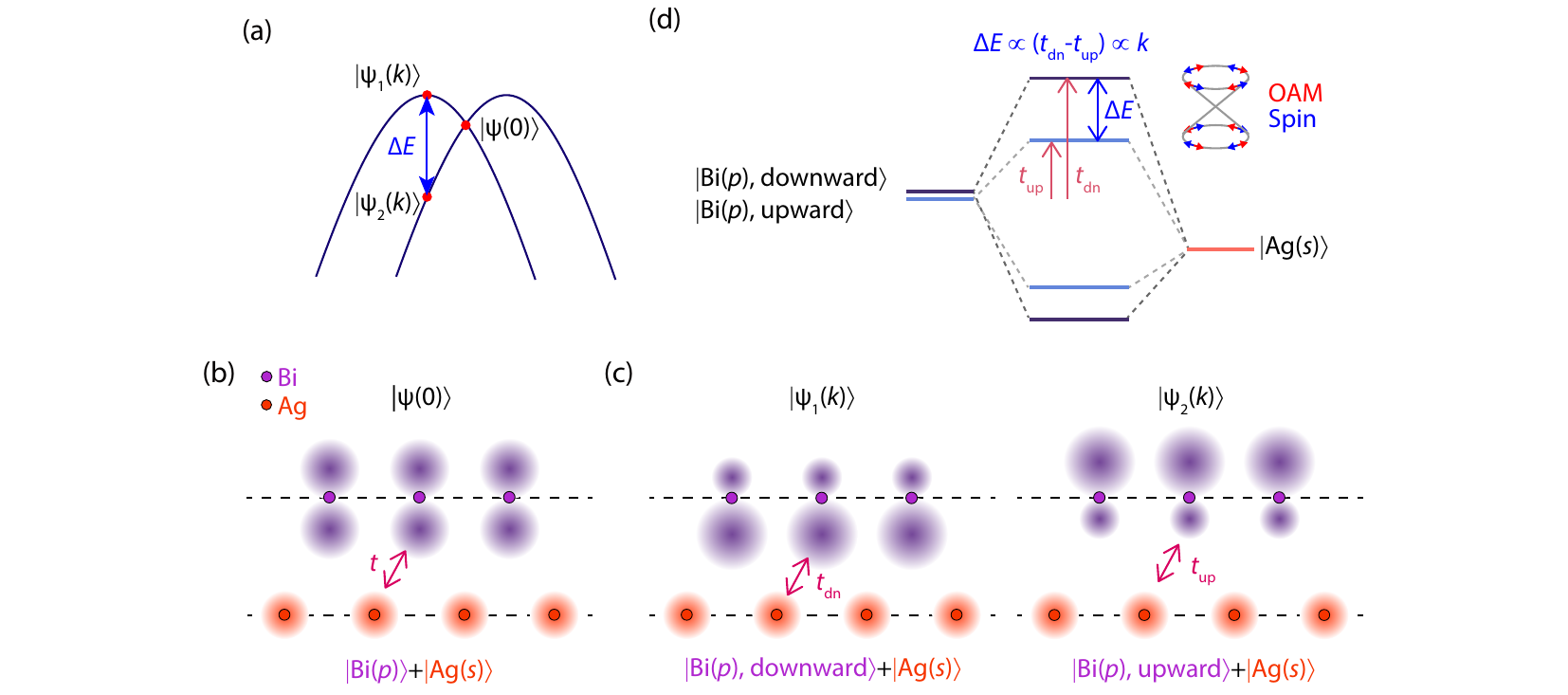}
\caption{\label{mechanism}
(a) Schematic band structure and wave functions.
(b),(c) Partial charge distributions at $k=0$ and $k\neq 0$.
(d) Orbital energy-level diagram for BiAg$_2$ and chiral spin/OAM of split bands.
}
\end{figure*}

In order to find the true mechanism of the giant RSS in BiAg$_2$, we use DFT and TB analysis to investigate the interaction between Bi and Ag. In Fig. \ref{along_k}(a), we plot the partial charge densities of the Rashba-split bands projected on the (110) plane passing through Ag-Bi-Ag atoms. In addition to the obvious Bi $p$-orbital feature, there is contribution from Ag orbitals which clearly indicates hybridization between Bi and Ag. Interestingly, there is a clear difference between the Ag contributions from the two Rashba-split bands; the partial charge density of the upper band has significant Ag contribution while the lower band has very little. As the momentum $k$ increases away from the $\Gamma$-point, the difference becomes more distinct. The difference is accompanied by an asymmetric charge distribution around Bi. We can see in Fig. \ref{along_k}(a) that the Bi states in the upper band spreads downward while they move upward in the lower band. This behavior is reminiscent of the $k$-dependent asymmetric charge distribution induced by the existence of local OAM \cite{Park11,Kim13,Hong15}.

Based on the linear combination of atomic orbitals (LCAO) coefficients from DFT calculation, we quantitatively investigate the composition of the bands, and present the difference in Ag $s$-orbital contributions ($\Delta$Ag($s$)) and $\Delta E$ for $\Delta z =$ 3, 4 and 5 \AA{} in Fig. \ref{along_k}(b). Both $\Delta E$ and $\Delta \text{Ag}(s)$ increase linearly with $k$, and increase as $\Delta z$ decreases. These results imply that OAM-induced asymmetric charge distribution around Bi results in different hybridization strength between Bi $p$- and Ag $s$-orbitals for the two split bands and that the difference in the hybridization dominates the energetics of RSS. We note that the two pairs of Bi $J\approx3/2$ bands located above the Bi $J\approx1/2$ bands have comparable or smaller RSS because the charge distributions of Bi $J\approx3/2$ states has dumbbell-like shape in the Bi plane, and is thus relatively less dispersive along the $z$-direction \cite{Hong15}. This results in a smaller overlap or hybridization between Bi and Ag orbitals.
 
To examine the correlation between $\Delta E$ and anisotropic hybridization strength in Rashba-split bands analytically, we consider a TB model for BiAg$_2$ monolayer (see Supplementary Information for detail). From TB analysis, we derive a Rashba Hamiltonian of the form
\begin{align}
H_R = E_\Gamma\sigma_0 + \alpha_R \mathbf{k} \times \boldsymbol\sigma\cdot \hat{z} \label{eq:rashba_ham}
\end{align}
around the $\Gamma$-point where $\alpha_R = 4\sqrt{3} V_{sp\sigma} A_s A_{xy}$ and $\sigma_i$ is the Pauli matrix. Here, $A_s$ and $A_{xy}$ are amplitudes of $s$- and $p_{x(y)}$-orbitals at the $\Gamma$-point. We assume that the lattice constant is equal to 1 for convenience. This leads to $\Delta E \approx 8\sqrt{3} |V_{sp\sigma}| A_s A_{xy}k$ and is well described by $\Delta\langle H_{sp} \rangle$, the difference between the averages of $s$-$p$ hopping energy between the two Rashba-split bands as shown in Fig. \ref{along_k}(c). Equation (\ref{eq:rashba_ham}) implies that the giant RSS stems from the large value of $V_{sp\sigma}$ ($\approx$1 eV) and the high $s$- and $p_{x(y)}$-orbital contents. ISB makes $A_s$ finite through the finite overlap between $s$- and $p_z$-orbitals, which allows the RSS. Then, the RSS is largely amplified by a large $V_{sp\sigma}$. These processes are simply characterized by $\Delta\langle H_{sp} \rangle$ in the long wavelength limit. 
We further show that $\Delta\langle H_{sp} \rangle$ is manifested as the physical quantity $\Delta\mathrm{Ag}(s)$ by the explicit formula
\begin{align}
\Delta E \approx \gamma \Delta\mathrm{Ag}(s)
\end{align}
where $\gamma$ is a constant evaluated from the eigenvalues and wave functions at the $\Gamma$-point.

Based on the DFT and TB analysis, we propose  a mechanism of the spin-dependent interatomic-hopping for RSS as schematically described in Fig. \ref{mechanism}. 
Since $k=0$ at the $\Gamma$-point (Fig. \ref{mechanism}(b)),  Bi $p$-orbital forms \textit{symmetric} charge distribution centered at the Bi atom. Thus, two spin bands are degenerate, as also required by the Kramer's theorem. Away from the $\Gamma$-point, asymmetric charge distribution around Bi develops as plotted in Fig. \ref{mechanism}(c) because of the local OAM in the $J\approx1/2$ state and finite crystal momentum $k$ \cite{Park11,Kim13,Hong15}. We denote those states by $|\text{Bi}(p), \text{downward}\rangle$ and $|\text{Bi}(p), \text{upward}\rangle$, depending on how the wave function extends. Note that charge densities of the two states extend in opposite directions because the OAM chiralities are opposite for the two states as seen in Fig. \ref{intro}(c). As a result, the two states have different hopping strength with Ag; $|\text{Bi}(p), \text{downward}\rangle$ and $|\text{Ag}(s)\rangle$ have a considerable overlap and a large hopping parameter $t_\text{dn}$, while $|\text{Bi}(p), \text{upward}\rangle$ and $|\text{Ag}(s)\rangle$ overlap less and thus have a smaller hopping parameter $t_\text{up}$. Consequently, the bonding and anti-bonding states of $|\text{Bi}(p), \text{downward}\rangle$ and $|\text{Ag}(s)\rangle$ have larger energy shifts than those of $|\text{Bi}(p), \text{upward}\rangle$ and $|\text{Ag}(s)\rangle$ as shown in Fig. \ref{mechanism}(d). Then, the energy difference between the two anti-bonding states determines the size of RSS. 
Note that the split energy is linear in $k$ because the charge asymmetry is proportional to $\mathbf{k} \times \mathbf{L}$.  

Our spin-dependent interatomic-hopping induced RSS model tells us that the direction of spin/OAM chiralities can be controlled by relative energy levels of the surface and substrate atoms. In the case of Bi/Ag(111), the energy level of the Bi $p$-orbital lies higher than that of Ag $s$-orbital. Therefore, Bi $p$-orbital mainly contributes to the anti-bonding states which show the giant RSS. If Bi $p$-orbital has lower energy than the substrate atomic orbital, Bi $p$-orbital would mostly contribute to the bonding states, and the bonding states will show giant RSS. In that case, $|\text{Bi}(p),\text{downward}\rangle$ will comprise the lower band, and spin/OAM chiralities will be opposite to those of Bi/Ag(111). Controlling the direction of chirality within our scheme will further back up our spin-dependent interatomic-hopping induced RSS model. 

An important aspect of our model is that it tells us three conditions to have a giant RSS. First, surface states should be localized within the length scale of the atomic distance to generate significantly asymmetric charge distribution upon ISB. Then, the surface atom must have a strong atomic spin-orbit coupling in order to have significant OAM ($J$ state) and thus develop $|\text{upward}\rangle$ and $|\text{downward}\rangle$ states under ISB. Finally, there must be proper overlap between surface and subsurface orbitals to maximize the energy difference between $t_\text{up}$ and $t_\text{dn}$.

%Acknowledgements
This research was supported by the National Research Foundation of Korea (NRF) grant funded by the Korea government (MSIP) (No. 2015R1A2A1A15051540). S.R.P. acknowledges the support from the NRF (2014R1A1A1002440). C.K. acknowledges the support by the Institute for Basic Science in Korea (Grant No. IBS-R009-G2). Calculation was supported in part by the Materials Simulation Center, a Penn-State MRSEC and MRI facility.

Note added -- We found a recent and independent work with a similar conclusion \cite{Sunko17}.

\end{document}